# What can Space Resources do for Astronomy and Planetary Science?

Martin Elvis [a,1]

[a] Harvard-Smithsonian Center for Astrophysics, 60 Garden St., Cambridge MA 02138, USA; melvis@cfa.harvard.edu

*revised: 22 July 2016*

## ABSTRACT

*The rapid cost growth of flagship space missions has created a crisis for astronomy and planetary science. We have hit the funding wall. For the past 3 decades scientists have not had to think much about how space technology would change within their planning horizon. However, this time around enormous improvements in space infrastructure capabilities and, especially, costs are likely on the 20-year gestation periods for large space telescopes. Commercial space will lower launch and spacecraft costs substantially, enable cost-effective on-orbit servicing, cheap lunar landers and "interplanetary cubesats" by the early 2020s. A doubling of flagship launch rates is not implausible. On a longer timescale it will enable large structures to be assembled and constructed in space. These developments will change how we plan and design missions.*

## 1. MORE LIGHT

Like the great poet and polymath Goethe, astronomers will be calling for "more light" on their deathbeds[2]. We are always seeking larger telescopes to collect the faint light arriving from the most distant stars, galaxies and quasars in the earliest times of the Universe; or else we are slicing up the light from bright stars exceedingly fine to look for signatures of small planets, other earths. There is no limit to what we crave. But we are in trouble. Our telescopes have grown in expense far faster than the economies they depend on. "*If something cannot go on forever, it will stop*" as Herbert Stein's Law states in economics [1]. What can we do to ensure that ever greater observatories lie ahead?

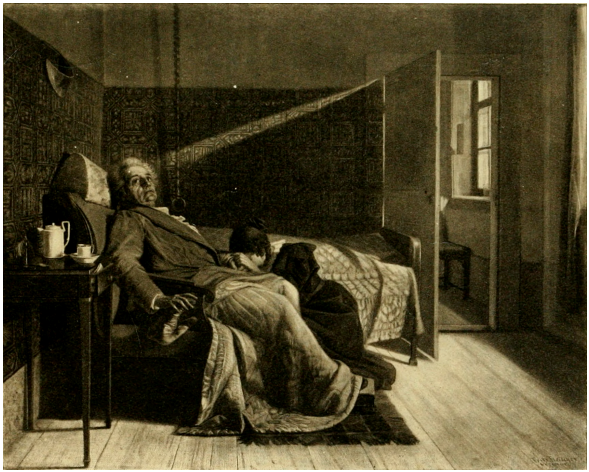

**Figure 1:** Johann Wolfgang von Goethe calls for more light on his deathbed (F. Fliescher; source: commons.wikimedia.org).

---

[1] Corresponding author.
[2] In Goethe's case this may well be mythical; for astronomers, not so much.

The new large space telescopes now being discussed will not launch for 15-25 years. On that timescale much is going to change that could help our field. In this paper I look at the rapid developments occurring in commercial space activities and examine how they could provide a way out of our dilemma.

These developments are numerous. In Figure 2 the timelines for major astronomy decisions to be made is compared with that for major developments anticipated for commercial space, including space resources. Clearly many relevant commercial space activities are set to happen before the next generation of major astronomy observatories are launched, or even begin their final design/build stages (phase C/D in NASA terminology). Planning to take advantage of these developments would seem advisable.

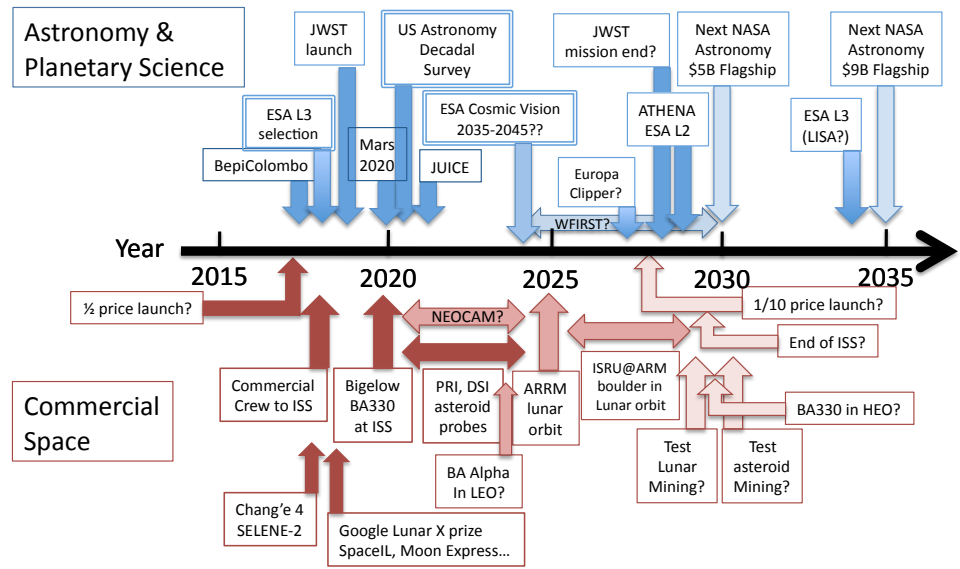

Figure 2: Comparison of timelines for astronomy and commercial space and space resources, including NASA activities that promote them. The "Next NASA Astronomy Flagship" markers assume a constant budget for large new astrophysics missions at NASA and no substantial external contributions. NEOCAM and ARRM are possible NASA missions, but they will enable commercial space ventures.

## 2. THE FUNDING WALL

Astronomers are already planning telescopes for the late 2020s. For example, the X-ray telescope ATHENA[3] has been selected as the second European Space Agency (ESA) Large mission (L2). It has an intended launch date of 2028. In the US the jostling for position to be given the #1 recommendation for large space missions in the 2020 "decadal study" has already begun. The astronomy management organization AURA has issued a report entitled "From Cosmic Birth to Living Earths"[4]. This report advocates for a telescope double the size of the *James Webb Space Telescope* (JWST) to take this #1 spot. This "High Definition Space Telescope" (HDST) would have advanced coronagraphic (starlight-suppressing) optics that would allow it to directly detect the light from a twin of the Earth around nearby stars. Their nominal launch date is 2035.

---

[3] http://sci.esa.int/cosmic-vision/54517-athena/
[4] http://www.hdstvision.org. See also my critique: arXiv:1509.07798, and the response: arXiv:1511.01144.

Why plan so far ahead? Are astronomers just keen on delayed gratification? There is a deeper reason for these long timescales.

2.1. GROWING AMBITIONS, GROWING COSTS

The reason we are making plans so far in advance is that our telescopes, in every band of the electromagnetic spectrum, have grown over the past few decades from small exploratory devices to Great Observatories. These flagship missions have enormous costs and take many years from conception to launch.

The prototypes for *Hubble* were space telescopes looking in ultraviolet light (which does not get through our atmosphere): the *Copernicus* Orbiting Astronomical Observatory, followed by the *International Ultraviolet Explorer* (IUE). These carried quite modest 80cm and 45cm diameter mirrors, respectively. *Hubble* has a 3-5 times larger mirror. JWST, billed as *Hubble's* successor, has a 6.5m telescope, almost 3 times larger still. If built, HDST would be another doubling. This tendency to grow in jumps of 3 or so in diameter, or about 10 in mirror area, is exponential growth. It is baked into our research programs, as discoveries just possible with the previous generation always need more light to discern what they are and how they work. Without an order of magnitude leap in at least one capability there is virtually no chance that your favorite flagship will fly. The problem is that, historically, larger telescopes cost more.

Cost growth can be tracked for any class of space missions. For many years my field was X-ray astronomy. The breakthrough satellite missions[5] came from NASA and were: *UHURU* (SAS-A, launched in 1970), the *Einstein Observatory* (HEAO-B, launched in 1978) and *Chandra* (AXAF, launched in 1999). Over the course of these 3 decades X-ray astronomy gained a factor of over 1 million in sensitivity. That is a truly huge advance, and is something that took optical astronomy about 200 years. The resulting impact on astrophysics was profound [2].

But the price was high. Figure 3 shows how the (inflation-corrected) cost of these missions increased by a factor of about 20 over 30 years. This is an exponential growth rate of 10% per year. The same plot for other wavelength bands would be much the same. Ian Crawford has shown that Mars landers have grown even faster, at about 15% per year [3]. Historical growth rates for the US GDP have been fairly steady at about 2% a year for the past century and more (1871 – 2001)[6]. Clearly, growth rates for astronomy that are four times faster than that of the economy are unsustainable.

Exponentially rising curves become all but vertical, so this mismatch of rates is often called "the funding wall" [4]. At some point the costs are more than a government can abide. Particle physics hit its funding wall in the US when the Superconducting

---

[5] With apologies to the many other fine missions that did sterling work, including Ariel V, which I used for my PhD thesis. Nonetheless, the factor 100 steps in sensitivity were those listed.

[6] The fastest decade of growth in US GDP since 1871 was in the 1941-1950 decade when the rate reached 3.87%. http://socialdemocracy21stcentury.blogspot.com/2012/09/us-real-per-capita-gdp-from-18702001.html

Super-Collider, already far along in construction in Texas, went over budget one too many times and was cancelled[7]. Is astronomy next?

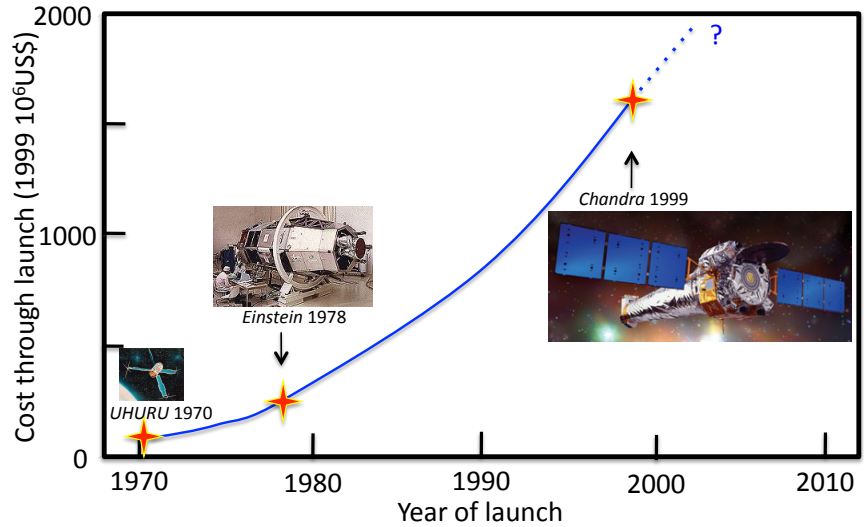

Figure 3: Cost growth of leading X-ray astronomy missions, 1970 - 1999. Cost is given in constant 1999 US dollars: $20M for *Uhuru* (1970$); $100M for *Einstein* (1978$); $1.6B for *Chandra* (1999$), H. Tananbaum, private communication. (Inflation corrections from US Bureau of Labor Statistics; URL: http://www.bls.gov/data/inflation_calculator.htm).

We may well be up against the funding wall right now. JWST is costing NASA almost $9B up through launch in 2018, with another $1B or so coming from ESA and Canada. Cost growth led to repeated cancellation threats but a de-scope in 2001 and a re-plan in 2011[8] averted this [5]. The HDST concept, AURA officials suggest[9], would cost about the same, though many outsiders are quietly skeptical that it would be so cheap. NASA currently has a budget of about $5B per decade for large new space telescopes[10]. Each JWST-class mission thus takes 100% of nearly 20 years of this funding line. So building HDST by 2035, about 10 years after the launch of the Wide-Field Infrared Survey Telescope (WFIRST) in the early 2020s[11], would require roughly doubling the available budget. A few billion more for such a major mission does not sound like an impossible target. But there is a catch.

2.2 THE NEED FOR PAN-SPECTRAL COVERAGE

The problem is that modern astrophysics depends on simultaneous access to the entire electromagnetic spectrum. Stars, galaxies, quasars, and even planets, blithely ignore the limitations of our technologies, emitting light across all wavelengths, from the radio and infrared to the optical, ultraviolet and X-rays. Once astronomers see a cosmic object across the spectrum problems that had seemed deeply mysterious are answered, like jigsaw pieces fitting together, as in the case of the "exploding galaxy" Messier 82 (Fig 4).

---

[7] Appell, D., (2013), http://www.scientificamerican.com/article/the-supercollider-that-never-was/
[8] http://www.space.com/12759-james-webb-space-telescope-nasa-cost-increase.html
[9] Calla Cofield, Space.com, http://www.space.com/29878-alien-life-search-hdst-space-telescope.html.
[10] http://files.aas.org/head2015_workshop/HEAD_2015_Paul_Hertz.pdf, integration of slide.14
[11] http://www.nasa.gov/content/goddard/qa-session-about-nasas-wfirst-mission

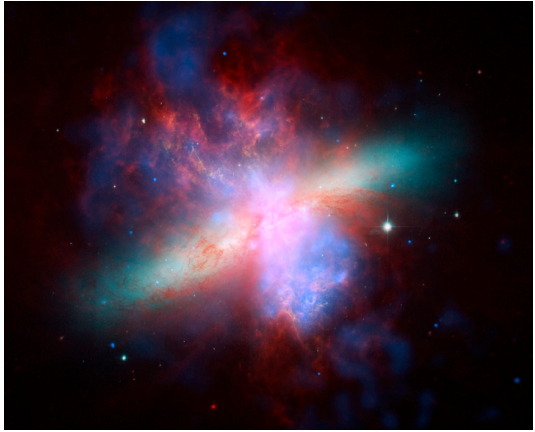

Figure 4: The 'exploding galaxy' Messier 82, seen in optical, infrared and X-ray light with *Hubble, Spitzer* and *Chandra*, NASA's three Great Observatories. No one observatory gives the whole picture; together the story is clear. (URL: http://chandra.harvard.edu/photo/2006/m82/m82_comp.jpg)

No one observatory can give the whole picture; together the story is clear: a giant burst of star formation in the center of the Messier 82 galaxy (*Hubble*, in green) forces a huge plume of gas heated to millions of degree (*Chandra*, in blue) out of the spiral disk, following the path of least resistance, with cool gas and dust following it around the edges. Here we are witnessing the process of young massive stars exploding as supernovae and sending their newly synthesized elements into space. It is from such materials that planets, and us, are formed. There are many such examples. They are the norm in 21$^{st}$ century astrophysics.

The synergy between these spectrum-spanning telescopes is surely a major reason that we are in a Golden Age of Astronomy.

We have been very fortunate, in fact, that this synergy has already lasted for 35 years, beginning around 1980 when we had *IRAS*[12], *IUE*[13] and *Einstein*[14] (and then *ROSAT*[15]) in quick succession spanning the infrared, ultraviolet, and X-rays respectively, right up to the present generation of Great Observatories: *Spitzer*, *Hubble* and *Chandra*. For virtually this entire time a discovery in one wavelength band could be followed up in the others just a year later. But the youngest of the three Great Observatories is 13 years old, and they cannot all be expected to last through the JWST era.

Even if they do survive they will not be matched to JWST's sensitivity. JWST will greatly surpass *Spitzer's* infrared capability after it launches in 2018. However, there is no chance that we will have matching ultraviolet or X-ray capabilities until 2030 or later. By then the JWST 5.5-10-year lifetime[16] will be over.

---

[12] IRAS 1983, January – November. (URL: http://www.jpl.nasa.gov/missions/infrared-astronomical-satellite-iras/). As IRAS was an all-sky survey later UV and X-ray discoveries could all still be followed up in the IRAS archive, despite its short lifetime.
[13] IUE 1978 – 1996 (URL: http://science.nasa.gov/missions/iue/)
[14] Einstein 1978 – 1981 (URL: https://www.cfa.harvard.edu/hea/hm/heaob.html)
[15] ROSAT 1990 – 1999 (URL: http://www.mpe.mpg.de/xray/wave/rosat/mission/rosat/index.php)

So following up a JWST discovery at another wavelength must wait, not a year, but a decade or more. The faint infrared glow from the earliest times in cosmic history (at $z>10$) will have to wait patiently for an X-ray measurement to tell us if it is powered by the first stars forming, or by an adolescent black hole having a growth spurt. When we do get that X-ray measurement we will find that no discovery with that new X-ray flagship can be followed up until JWST too is replaced.

That kind of long delay will slow progress in the field to a crawl compared to today. Not the least effect will be that the dimmed excitement will put off talented young scientists from entering astrophysics.

Planetary science is similarly hindered by high mission costs. The 2011 Planetary Science Decadal report "Vision and Voyages" recommended three large-class missions, one each to Mars, Europa and Uranus[17]. But only one can be done per decade[18], making this a 30-year program for even the top handful of priorities. The Mars mission was chosen[19], and it is only the first half of a Mars sample return program. Yet the more we learn of Solar System worlds the more we realize what we don't know. The 2015 New Horizons encounter with Pluto demonstrated this once again[20]. This is not exploring the Solar System at scale. We need to do far more.

2.3 THE CRISIS

Cost growth has created a crisis for both astronomy and planetary science. As noted in Section 2.1, a single new flagship following JWST and WFIRST might be as expensive as its ~$9B budget and yet launch relatively soon by obtaining a doubling of NASA's large astronomy missions budget. It is presently inconceivable though that we could obtain enough funding to launch four such flagships in a single 20 year span, maintaining spectral coverage. Some observing "window" will have to be let go. And that is ignoring the new field of gravitational wave astronomy [6]. Quadrupling or quintupling NASA's budget for large astrophysics missions would be a huge stretch. On this path we get our "Greater Observatories" sequentially.

And what would the longer-term prospect be even if we did quadruple the budget? What will we do when the science pushes us to even larger telescopes, or more ambitious ones? For example, an optical interferometer that can resolve the inner workings of quasars, and image continents and oceans on twins of the Earth orbiting other stars is an idea that has already been studied. It was called the "Terrestrial Planet Imager - Interferometer" (Figure 5), but it has been set aside since 2008[21]. We know how to build it conceptually, but it exceeds today's construction capabilities in space, and we can't afford it at current prices.

---

[16] Limited by station-keeping hydrazine fuel supply. A few year extention may be possible. http://jwst.nasa.gov/faq_scientists.html#lifetime
[17] Vision and Voyages for Planetary Science in the Decade 2013-2022 (2011) http://www.nap.edu/catalog/13117/vision-and-voyages-for-planetary-science-in-the-decade-2013-2022
[18] Although the Europa mission was revived in a cleverly designed cheaper mission, *Europa Clipper*, it now seems unlikely to launch before the "late 2020s" (http://www.space.com/31887-nasa-europa-mission-launch-late-2020s.html); the once per decade cadence remains in place.
[19] in modified form as *Mars 2020* (http://mars.nasa.gov/mars2020/mission/overview/)
[20] New Horizons web site: http://pluto.jhuapl.edu

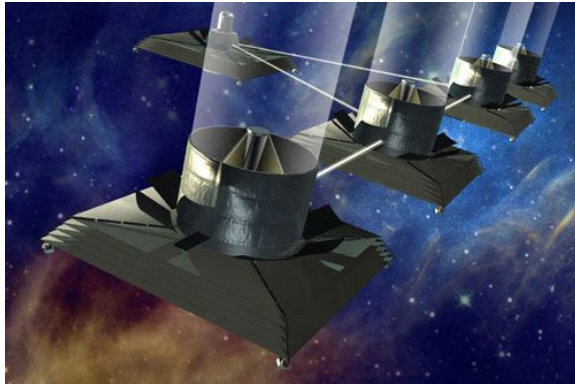
Figure 5: Terrestrial Planet Imager - Interferometer. (Source: NASA)

If we do not think creatively about this crisis we will all be competing for the one place at the shrinking water hole. This is an unattractive future for a great endeavor. It is one that effectively closes many of the observing windows we have opened in the past 50 years of the space age. How can we envisage a greater, open-ended, future for astrophysics and planetary science?

The answer, I argue, lies in the commercial harnessing of space resources. Alan Stern, PI of the New Horizons mission to Pluto argues similarly [7]. Astronomers are planning on 20-year timescales, but are not yet fully aware of the major changes coming to commercial space activities in the next two decades. Most of us do know about SpaceX and its re-useable rockets, but perhaps not so much beyond that.

In the next section I look at the commercial space developments in the near-future to around 2020 and examine the effects they will have on astrophysics and planetary science planning. On this timescale the developments are quite predictable. In the following section I look out to the developments happening until around 2030, when the next flagship missions will launch. These are more speculative, but also more profound.

## 3. THE NEAR FUTURE TO 2020

To date the only space resource used commercially has been non-material. Location is foremost. The few kilometer thick band of geostationary Earth orbit (GEO) has long been valuable real estate[22]. Low Earth orbit (LEO) is newly prized as a location for constellations of for-profit Earth-observing satellites[23] but is not yet in short supply.

---

[21] https://exep.jpl.nasa.gov/TPF-I/tpf-I_what_is.cfm
[22] http://history.nasa.gov/satcomhistory.html
[23] http://www.thespacereview.com/article/2716/1

Micro-gravity is another non-material space resource for which new commercial uses are imminent.

So what can commercial space do for astronomy and planetary science? Potentially, it can do a lot. All commercial activities feel intense pressure to bring down costs in order to maximize profits, and commercial space is no different. Astronomy and planetary science can exploit these savings. How large could these savings be?

A rule of thumb is that launch and the science payload each consume about a quarter of a mission budget (up to launch), while the spacecraft consumes about half. Table 1 quantifies the potential gains on each element of a mission for factors 2-3 cuts in launch and spacecraft cost. (As discussed in Section 3.3, science payload savings are capped at 50%.) Not too surprisingly, for cost reductions of factors 2-3, similarly more missions could be flown.

Table 1: Potential Effects of Launch, Spacecraft and Science Payload Cost Reductions[a]

|  | Earth Orbit | | |
|---|---|---|---|
|  | Now | ÷2 | ÷3 |
| Launch | 25 | 13 | 9 |
| Spacecraft | 50 | 25 | 17 |
| Science Payload[b] | 25 | 13 | 13 |
| TOTAL | 100 | 51 | 39 |
| Number of missions | 1 | 2 | 3 |

a. All numbers rounded to avoid fractions; b. Limited to factor 2 savings, as optics and sensors must remain state of the art (section 3.3).

There are five near-term areas where commercial space would effect changes: (1) launch cost reduction; (2) cheaper spacecraft; (3) cheaper payloads; (4) cheap, quick on-orbit testing; and (5) cost-effective servicing. I consider each of them in turn in the following sections.

3.1 REDUCED LAUNCH COST

The dramatic soft-landing of a Falcon-9 first stage on the barge *Of Course I Still Love You* in the Atlantic Ocean on the day this talk was given (8 April 2016)[24] made clear that re-useable rockets are rapidly coming of age. Stimulated by SpaceX, both of the traditional launch providers, Airbus and United Launch Alliance (ULA), announced their own variations on the concept in 2015[25]. In addition Blue Origin have demonstrated their

---

[24] http://www.cbsnews.com/news/falcon-9-boosts-cargo-ship-to-orbit-sticks-ocean-landing/
[25] Airbus "Adeline" concept, June 2015: http://www.space.com/29620-airbus-adeline-reusable-rocket-space-tug.html; ULA Vulcan concept, 13 April 2015: http://spaceflightnow.com/2015/04/13/ula-unveils-its-future-with-the-vulcan-rocket-family/

New Shepherd re-useable first stage for sub-orbital flight and have made clear that an orbital version is planned[26].

The launch industry is on track to reduce launch costs substantially with these re-useable rockets. A factor 30% - 50% cut in launch cost is possible for the first Falcon-9R commercial re-useable launch, probably in 2017[27]. As fuel accounts for only about ½ % of launch costs, factors of 10 or more price reductions are conceivable. For now the timescale to achieve such large factors is unclear. If SpaceX development goes as fast as they say, the late-2020s is a reasonable guess at when factors this large might happen.

If launch were to be 10 times less expensive this would provide a useful saving for a given mission, as launch typically amounts to ~25% of mission costs through launch. It would also improve the cadence of missions, building 4 in the time it would now take to fund 3. But launch savings alone are not enough to allow two major observatories to be built simultaneously. Even if launch were free it would not alone be enough to double the number of missions.

It is not cheaper launches by themselves that will change the economics of space missions, but their effect in enabling low cost spacecraft, as I discuss below.

3.2 CHEAPER SPACECRAFT

For most major astronomical observatories in space, the spacecraft is about half the cost. Cutting spacecraft cost by a factor 2, along with launch costs by a factor 3, would roughly halve the cost of a flagship mission. This would radically alter flagship mission decisions, by doubling the number that could be launched each decade. This raises the question "Why are spacecraft so expensive?"

We usually use the term "aerospace engineering". This gives the impression that designing aircraft and spacecraft are a single discipline. But in fact space engineering is driven by different considerations than aircraft engineering, and so is a different discipline. Above all, for space, mass matters. Mass is a consideration for aircraft design, of course. But minimizing mass in spacecraft is paramount, thanks to the ineluctable rocket equation that demands exponentially more propellant for linear increases in payload mass.

This focus on mass has consequences. On a spacecraft every kilogram, even every 100 g, has to be carefully argued for. This pushes spacecraft design toward designs using exotic components and relatively small performance margins, and that leads to rigorous design and test regimes. Each spacecraft, with the exception of communication satellites (comsats), is typically a bespoke tailored design. This approach makes spacecraft extremely expensive. For example, I sat in one meeting in which several well-paid engineers flew in to spend an afternoon learning the basics about a 100 g sensor that cost about $1000, plus a meter or two of cabling. That one meeting cost more than the sensor,

---

[26] https://www.blueorigin.com/technology
[27] spacenews.com/spacex-says-reusable-stage-could-cut-prices-by-30-plans-first-falcon-heavy-in-november/; http://selenianboondocks.com/2016/04/spacex-amateur-business-case-study/; http://spacenews.com/spacexs-reusable-falcon-9-what-are-the-real-cost-savings-for-customers/

and it was the first of many on just that one small detail of the whole payload. Until we can relax the mass design constraint we will have highly costly spacecraft.

Mass is the driver for spacecraft design because of the daunting cost of getting to orbit, presently about $10k/kg to LEO. Reduced launch costs will have a knock-on effect, as the entire space engineering discipline will gain much greater flexibility about its design drivers. Using (relatively) massive spacecraft with large performance margins is now unaffordable. Once launch costs are cut by factors of a few then a "fat-sat" approach will be a natural path to lower cost spacecraft [8]. Sturdy spacecraft structures, multiply redundant electronics, electrically shielded cabling, larger solar arrays allowing the use of higher power consumption electronics are examples of the many ways in which spacecraft design constraints would be eased, and design/test cycles simplified. A Boeing study, cited in [8], indicates that a 50% growth in spacecraft mass could lead to a factor 3 reduction in spacecraft cost. If over-capable spacecraft are acceptable then they can be "off-the-rack", like cubesats, rather than bespoke, and can reap the cost advantages of batch production.

This idea has a long history [8], but until now has lacked the necessary factor of about 3 reduction in launch costs to make the transition happen. By the mid-2020s this barrier should be gone (Section 3.1).

3.3 CHEAPER SCIENCE PAYLOADS?

The remaining big ticket item for a mission is the scientific payload. Telescope mirrors and other optics, sensors and pre-amp electronics typically all push the envelope of technology. They have to, in order to produce a major, world-beating, advance. This part of payloads is not likely to get cheaper.

Science payloads, though, include many other components too. These include: structures, power supplies, thermal control, down-stream electronics and data processing. These components could benefit from the same design criteria relaxation as the spacecraft itself. Higher mass and power could be tolerated for them. What factor cost reduction this relaxation might enable has not yet, to my knowledge, been studied.

3.4 PASSENGER FLIGHTS TO ORBIT

The commercial launch of passengers into space is progressing fast. In the near future, by 2018, both SpaceX and Boeing will have operational spacecraft, *Dragon-2*[28] and *CST-100 Starliner*[29] respectively, transporting up to seven crew at a time to the International Space Station (ISS) as part of NASA's Commercial Crew Program[30].

Sub-orbital passenger spacecraft are also being developed as precursors to orbital capabilities. Although Virgin Galactic has been the highly visible face of sub-orbital space tourism for a decade, they do not seem to have a direct route to orbital flights. Instead other companies have clear development paths to orbit. Blue Origin, funded by Amazon founder Jeff Bezos, had a spectacular public debut in November 2015 when they separately soft-landed both their (empty) crew capsule and their single stage rocket back at the launch site. They are clear that they view sub-orbital as a natural path to orbital flights. Their commercial orientation is clear in the video they released[31]. It touted "the largest windows in space", inviting comparison with Virgin Galactic. Advertising the final frontier may seem gauche, but it will pay the bills. Less splashy is XCOR[32]. Their small Lynx spacecraft will begin, like Virgin Galactic, as a sub-orbital vehicle, but is designed to scale up to an orbital capability[33]. Sierra Nevada may be a fourth orbital contender. For now their Dream Chaser is only slated for cargo delivery to orbit[34], but this is actually a downscaling of their original personnel carrier.

---

[28] http://www.spacex.com/news/2014/05/30/dragon-v2-spacexs-next-generation-manned-spacecraft
[29] http://www.boeing.com/space/crew-space-transportation-100-vehicle/
[30] https://www.nasa.gov/exploration/commercial/crew/index.html

If the market for orbital joy rides and for trips to commercial space stations (section 3.7) proves to be robust, there will be many companies competing for passengers by the early 2020s. XCOR is aiming for a ticket price of $1M to orbit. That is a price point that opens up many more customers, and can plausibly be matched by the other providers, if launch prices come down by an order of magnitude as SpaceX have often claimed[35].

3.5 INSTRUMENT TESTING ON-ORBIT

The advent of commercial passenger transport will likely include opportunities for instruments as piggyback payloads. The *Dragon-2* "trunk" space is one location[36]. The *Lynx* is designed to carry external and internal payloads as alternatives to a passenger[37].

Already cubesats are starting to enable hardware tests to high technical readiness levels (TRL-7[38]) at modest cost. Instruments that are too large for 3U or 6U cubesat formats, or that need more power than they can supply, could be tested in the space environment on passenger flights, although not in the integrated mission operations conditions needed for the top readiness level, TRL-9. A capability like this could accelerate the development, improve the reliability, and reduce the cost, of science payloads for large missions at modest cost. This could allow more advanced instruments to be flown at an earlier date.

3.6 COST EFFECTIVE SERVICING IN LEO

If space tourism extends to spacewalks (Extra Vehicular Activity or EVAs in NASA-speak) then cost-effective on-orbit servicing in LEO will become feasible. The scientific value of on-orbit servicing of spacecraft is well-established. The five servicing missions for *Hubble* were dramatic technical successes[39]. The first, SM1, was a true *tour*

---

[31] https://www.youtube.com/watch?v=9pillaOxGCo
[32] http://www.xcor.com
[33] http://spacenews.com/suborbital-vehicle-developers-looking-ahead-to-orbital-systems/
[34] beginning in 2019. http://spectrum.ieee.org/tech-talk/aerospace/space-flight/nasa-contracts-dream-chaser-shuttle-for-space-station-resupply
[35] http://arstechnica.com/science/2015/12/by-making-a-historic-landing-spacex-launches-new-age-of-spaceflight/
[36] http://www.spacex.com/news/2013/03/26/dragon-trunk
[37] external: http://science.xcor.com/payloads/external-payloads/
[38] https://www.nasa.gov/directorates/heo/scan/engineering/technology/txt_accordion1.html
[39] http://www.nasa.gov/mission_pages/hubble/servicing/index.html

*de force* by both the ground teams and the astronauts. SM1 corrected the aberration in the primary mirror, rescuing the mission. Subsequent missions replaced all the original instruments with new versions that were orders of magnitude better in some parameters, and extended the life of the mission by replacing key components that had failed (gyros and batteries) and by boosting *Hubble* to a higher orbit.

However, the cost of the *Hubble* servicing missions[40] was sufficiently high that the two successor Great Observatories were placed in orbits that could not be serviced. In the case of *Chandra* a highly elliptical orbit was deliberately chosen to prevent servicing, in order to keep down the post-launch run-out costs of the program[41] and avoid cancellation. This orbit turns out to have science advantages, but that was not the initial impetus for choosing it. *Spitzer* was put into a novel Solar drift-away orbit. This was chosen to reduce the heat load on the on-board cryogen so as to allow less to be carried aboard while keeping a 5-year cold mission lifetime. This change greatly reduced the mission mass and cost through launch, enabling the mission to be approved [9]. It also served to prevent any possibility of servicing, and so kept run-out costs down, as for *Chandra*.

With low cost human access to LEO, servicing can return to being a normal part of missions in accessible orbits. Replacing failed systems can allow higher risk profiles and so lower costs. A lost mission, such as *Hitomi*[42], might be revived at a fraction of the cost of a new mission, and a repair would likely be much faster to implement. On-orbit servicing could also encourage agencies to risk choosing more cutting edge instruments to be installed in telescope focal planes. An instrument failure then becomes a nuisance, not a mission-ending event. Astronauts valued the Hubble servicing missions highly. "*As an astronomer and an astronaut, putting my two hands on Hubble in space was one of the greatest thrills of my life*", says Jeffrey Hoffman[43] who took part in SM-1. If servicing missions return, they would be motivationally valuable to the astronaut corps.

3.7 COMMERCIAL SPACE STATIONS IN LEO

Timed to follow the introduction of commercial passenger flights to orbit, Bigelow Aerospace, now in partnership with ULA[44], intends to launch its first BA330, 330 cu. meter, privately built and funded space station in 2020 to attach to the ISS. The primary resource provided there will be micro-gravity. Their idea is to rent out volume not only to the major space agencies, but also to large corporations, foundations and research organizations, as well as to countries wanting a cheaper route to a space program. If the ISS reaches its end-of-mission in 2024, as could happen, the BA330 could then be detached and operated independently.

---

[40] About $1.2B/mission (2010 dollars) in operating costs, discounting development costs: http://www.space.com/11358-nasa-space-shuttle-program-cost-30-years.html
[41] Leary, W.E., 1998, http://www.nytimes.com/1998/03/31/science/telescope-will-offer-x-ray-view-of-cosmos.html?pagewanted=2
[42] *Hitomi*: http://global.jaxa.jp/projects/sat/astro_h/;
[43] private communication, 20 April 2016.
[44] http://spacenews.com/ula-and-bigelow-announce-partnership-for-launching-commercial-space-stations/

An advantage of a free-flying Bigelow space station will be a major shrinking of the 2½ year training program for ISS-bound astronauts [10]. Such a long training for skilled bench scientists would be prohibitive for commercial companies. All the BA330 equipment will come from a single supplier and not, as on the ISS, from US, Russian, European and Japanese suppliers. This will simplify the training of the scientists who take passenger flights to orbit. Not being required to learn Russian or to spend long periods in Russia is another advantage.

We know how much customers would pay. Bigelow Aerospace advertises rates of $25M/110 cubic meters for 60 days[45]. That translates into a substantial $450M/year in revenue for one BA330 module. Their follow-on Space Station Alpha concept[46] has two BA330 modules and so would generate almost $1B/year in revenue, assuming customers materialize. NASA's Bigelow Expandable Activity Module on ISS[47] (BEAM, 2016-2018) is a first test of their technology with astronauts involved.

The main customers for a private space station may well be large biotech companies. Astronomers, including myself, have tended to think that no interesting laboratory research has been done at the ISS. We are not well-informed. In fact, gene expression can be radically altered in bacteria and plants grown in space [11] and the changes can be passed down to subsequent generations, even though they are back in Earth gravity. Biotech is a $270B/year research-intensive industry that is growing at 12% per year[48]. By 2020 biotech will be a half trillion dollar industry. If gene expression changes in low gravity are common, then it is highly plausible that biotech companies could put 0.2% of their revenue into space-based research, fully booking Alpha's facilties.

By 2020 then, when US astronomers sit down to contemplate their future in the next decadal study, it is plausible that a commercial laboratory in space will have begun, and by 2025 a free-flying commercial space station may well be operating.

The benefits of private research laboratories for astronomy and planetary science are less clear than the other developments. The BA330 does not have external trusses or platforms that could hold zenith, sky-pointing, pointing telescopes. (Nor could it accommodate nadir, Earth-pointed, instruments.) A specialized BA330 fitted out for cosmic ray detection is a possibility, but this is a relatively narrow application.

The main benefits would probably be the increased traffic to LEO that it would generate. If 2 months and 110 $m^3$ is the standard rental, then there would be 6 teams on-board Alpha with changeovers 6 times/year. If a team is 2 scientists (e.g. to work in shifts), then 12 flights/year using CST-100/Dragon-2 on Atlas-V/Falcon-9 class launchers would be needed to ferry up the scientists. That implies monthly crewed flights. While this is a minor increment to the current ~50 flights/year of this vehicle class, it is a factor

---

[45] http://www.nasaspaceflight.com/2014/02/affordable-habitats-more-buck-rogers-less-money-bigelow/
[46] http://www.space.com/19291-inflatable-alpha-station-bigelow-aerospace.html
[47] Launched on the day this talk was given, 8 April 2016.
http://www.nasa.gov/mission_pages/station/research/experiments/1804.html
[48] http://www.grandviewresearch.com/industry-analysis/biotechnology-market

4 increase over the present rate of crewed flights of ~4 flights/year[49]. Together with orbital tourism, these flights would make on-orbit testing (section 3.5) and servicing (section 3.6) more routine.

3.6 PRIVATE LUNAR LANDERS

Several countries are already planning or executing lunar lander programs. China put *Chang'e 3* and its *Yutu* rover onto the Moon in 2013, demonstrating technological readiness. The follow-up *Chang'e 4* mission is planned to land on the lunar far side[50]. Japan's JAXA space agency should fly the SELENE-2 lunar lander [12] in 2017[51].

On the non-state actor side, the Google Lunar X-prize[52] has inspired academic groups, non-profits and private companies alike to design cheap lunar landers. The $30 M Google Lunar X-Prize will close out at the end of 2017. As of writing two teams have launches booked in 2017: SpaceIL with Space X, and Moon Express with Rocket Lab USA. (Rocket Lab's *Electron* rocket had not flown as of writing[53].) Moon Express has 3 dedicated flights reserved at a reported $5M per flight[54]. Moon Express plans to attempt lunar sample return by the third flight ~2020. Moon Express in particular has a clear commercial focus for its MX-1 spacecraft/lander. Prices are claimed to be "less than $50 M"[55].

For astronomy these small landers offer little at first. They may enable prototype far-side very low frequency (<30 MHz) radio telescopes.

For planetary science the advantages of these small landers are clear. There have only been a few lunar sites visited with scientific equipment. These were all chosen decades ago, long before the Moon was mapped in detail. Today we have kilometer scale or better measurements of the Moon at many wavelengths, plus detailed gravity maps [13]. These maps show that the Moon is highly non-uniform and they highlight very many places that planetary scientists would like to land, study, and from which they could return samples. The Lunar Geophysical Network, for example, proposed as a NASA New Frontiers mission may be enabled by this new capability. Well-chosen locations for ground truth will also multiply the value of the global surveys from lunar orbit.

3.7 ASTEROID PROSPECTING

---

[49] for 2015: https://en.wikipedia.org/wiki/2015_in_spaceflight
[50] Space Daily, 22 May 2015. 'China Plans First Ever Landing On The Lunar Far Side'. http://www.spacedaily.com/reports/China_Plans_First_Ever_Landing_on_the_Dark_Side_of_the_Moon_999.html
[51] http://www.cnn.com/2015/04/23/tech/japan-moon-lander-planned/ Accessed 15/07/2015.
[52] http://lunar.xprize.org Accessed 15/07/2015.
[53] A mid-2016 launch was planned as of writing (http://spacenews.com/rocket-lab-plans-to-begin-launches-mid-year/)
[54] http://www.nzherald.co.nz/nz/news/article.cfm?c_id=1&objectid=11300831
[55] Richards, R., presentation to European Space Policy Institute, Vienna, Austria, 2016, April 16: https://www.youtube.com/watch?v=ealvx3Gx1Mw

Asteroid mining companies have been in the news quite a bit. There are two main ones: Planetary Resources Inc[56] and Deep Space Industries[57]. Both plan to mine water from near-Earth asteroids as their first "ore", and both expect to send out their first small experimental interplanetary spacecraft to asteroids within 5 years. Water ore bodies among the asteroids may be relatively hard to find [14] requiring large numbers of probes, each one of which must then be cheap in order to close the business case.

The interplanetary cubesat-class spacecraft being developed for asteroid prospecting[58] are intended to cost around $5M each, an extraordinarily low price. Their instrument-carrying capacity is small in mass and volume (~1kg, ~1U, i.e.10×10×10cm). Nonetheless, they offer the opportunity to field large numbers of highly targeted planetary missions to the Inner Solar System, and to do so quickly.

By 2020 the companies may be ready to sell copies of these spacecraft for science missions. For planetary science, as there are 24 different sub-types of asteroids [15], a fleet of small missions, with very limited instrumentation on each, could be considered to explore them all. The cost of this fleet would likely be less than that of a single NASA New Frontiers mission (~$700M for New Horizons[59]) and so would be something of a paradigm shift from the current approach of sending one heavily instrumented spacecraft to a single destination (e.g. OSIRIS-REx [16]).

3.8 THE STATE OF PLAY IN 2020

It is useful to bring together all the developments considered so far along with their implications for astronomy and planetary science. Table 2 summarizes the situation.

Table 2: Commercial Space activities effects on Astronomy and Planetary Science to 2020

| COMMERCIAL SPACE ACTIVITY | EFFECT ON COST | VALUE TO ASTRONOMY, PLANETARY SCIENCE |
|---|---|---|
| Lower cost launch (≥2X) | Cheaper Spacecraft. (factor 2-3?, "off-the-rack") | "2 for 1" missions, both astronomy and planetary science |
| | Cheaper science payloads (factor 2?) | |
| Passenger flights to LEO | Cheap High TRL tests | Cutting edge instruments |
| | Cost-effective servicing in LEO | Higher risk sub-systems in LEO |
| Private Space Stations | More passenger traffic | Indirect via more passenger flights |
| Private Lunar Landers | Cheaper access to lunar surface | Many expeditions to scientifically interesting sites |
| Asteroid Prospecting | Cheaper access to inner Solar System | Many specialized missions to all near-Earth asteroid types |

---

[56] http://www.planetaryresources.com
[57] https://deepspaceindustries.com
[58] "*Arkyds*" by Planetary Resources (http://www.planetaryresources.com/arkyd/); "*Fireflys*" by Deep Space Industries (https://deepspaceindustries.com/prospecting/).
[59] p.4 of https://www.nasa.gov/pdf/139889main_PressKit12_05.pdf

These are mostly highly probable developments. Together they will significantly change the landscape in which astronomers and planetary scientists plan their next steps. Cheaper launch and spacecraft, with on-orbit servicing will likely be the first significant gains for science to result from commercial space. In this way location and micro-gravity can be helpful space resources even before material space resources are exploited. Other changes will take longer, as we shall see next.

4. LONGER TERM TO 2030: MATERIAL SPACE RESOURCES BEGIN TO MATTER

To properly match the astronomers' planning horizon we need to look ahead by another decade, to 2030. Plans on that timescale are necessarily more speculative, yet there are enough indicators that we can make reasonable extrapolations. The big change is the move away from LEO as the use of material space resources begins. The benefits of the beginning of lunar and asteroid mining for astronomy and planetary science now begin to be large. These benefits lie in three areas: (1) on-orbit assembly; (2) on-orbit construction; (3) rapid Solar System transit times. First I describe the potential developments, then I cover each of the three areas in turn.

4.1 A PRIVATE SPACE STATION IN HIGH ORBIT

It is quite plausible that by 2030 a commercial space station could have been placed in a high orbit, perhaps at the Earth-Moon L1 Lagrange point (EM-L1). The commercial motivation for doing so is to attain true micro-gravity. A government laboratory would have the same imperative and both may be deployed.

Atmospheric drag and gravity gradient torques limit ISS and other LEO stations to milli-gravity [17]. Sensitive experiments will benefit from the quieter environment at a Lagrange point, far from atmospheric drag and at zero gravity gradient. To maintain micro-gravity astronauts could not live on such a station. They could visit to tend the experiments, fix hardware, resupply consumables, or retrieve products.

A deep space, e.g. EM-L1, space station could act as a catalyst for commercial space mining operations. The radiation environment outside the radiation belts and specifically at EM-L1 is much worse than in LEO. Genetic experiments must be sensitive to radiation dosage. Hence to control the experimental conditions biotech companies will require heavy shielding. A naïve calculation making the mass of shielding equal to the mass of the Earth's atmosphere at sea level suggests of order 10 tons of shielding per square meter. Hence thousands of tons of shielding would be needed for any reasonable sized laboratory. Water makes good radiation shielding and this could become a usefully large market for resources from space.

Space-supplied shielding is natural to consider as, in such high orbits, it is energetically much cheaper to bring the raw materials for these activities from the Moon or from some near-Earth asteroids. It may even be economically cheaper: even if launch costs come down by a factor 10 the price is still $1M/ton from Earth to LEO[60]. To reach high orbits would multiply the cost of supplies from Earth by another factor 2 – 3. At

---

[60] Metric units used throughout. 1mt = 1000 kg = 2204 lb = 1.1 US ton.

these rates adding a thousand tons of shielding makes even a cheap space station into a billion dollar plus investment again all too quickly. That's a daunting figure for a commercial company. If space-derived materials could be supplied for, say, a quarter the price, then there would be an opening for both space miners and deep space research stations to flourish. That is a big if, of course.

The value of an EM-L1 station for both astronomy and planetary science is quite major. Both cargo and passenger transport from LEO to EM-L1 will be needed. This will enable deep space servicing of astronomical telescopes. The Earth-Sun L2 point (ES-L2) is 1.5 million kilometers from Earth, about 4 Lunar distances. ES-L2 is a favorite location for sensitive astronomical observatories. *Herschel, Planck*, and *JWST* all use this orbit. With a delta-v of 0.4 km s$^{-1}$ [18] it is energetically easy to reach ES-L2 from EM-L1. Planetary Science will gain the ability to fuel interplanetary probes at the top of the Earth's gravity well, enabling large payloads to be delivered.

Will suppliers of water, or other shielding material, from space resources be ready to meet the demand? Next I look at the feasible state of development of asteroid and lunar resources by 2030.

4.2 ASTEROID RESOURCES

On the 2030 timescale several developments will combine to make true asteroid mining plausible.

(1) By 2025 the two asteroid mining companies should be sending out fleets of their cheap "interplanetary cubesat" spacecraft to asteroids that look like good mining prospects (see section 3.7). By 2030 these proximity probes should have found several ore-bearing, i.e. profitable, asteroids. [The knowledge of which ones they are will, of course, be intellectual property (IP) belonging to the companies. Science will not benefit directly from that knowledge.] So the first robotic mining forays can begin around 2030.

(2) High power solar electric propulsion (SEP) will also have been demonstrated by 2025. This will enable interplanetary orbits for larger spacecraft. NASA's Asteroid Redirect Mission (ARM) will use 40 kW SEP to move a 7 - 10 ton spacecraft, including the xenon propellant [19] out to a near-Earth asteroid. This class of SEP is sufficient to return substantial material mined from an asteroid. The ARM spacecraft will take ~2 years to return a ~10 ton boulder[61] to a lunar orbit in ~2021. If more accessible asteroids can be found this mass can go up to 1000 tons [20].

(3) Astronomical surveys for near-Earth asteroids (mainly supported by NASA) will have found vastly more objects by 2030. The Catalina Sky Survey[62] in Arizona and Pan-STARRS[63] in Hawai'i are progressing apace, at a rate of ~1500/year in 2015, to a current total of ~14,000. This rate will accelerate in 2016 once Pan-STARRS-2 and ATLAS[64] come on line. A naïve projection is that 30,000 near-Earth objects will be known by 2030 from these surveys. When the Large Synoptic Survey Telescope (LSST) comes on-line in the early 2020s it will find many more near-Earth asteroids [21]. Many of these near-Earth asteroids will be small and must have orbits bringing them especially close to the Earth just to be detected. This could provide a supply of lower delta-v asteroids in the 1000 ton class that could be excellent mining targets, especially for water. If NASA approves the JPL NEOCAM[65] mission then from about 2021 to 2023 a thermal infrared survey will catalog some 100,000 near-Earth objects, and get good sizes for them.

---

[61] Chodas, P. (2015) http://www.lpi.usra.edu/sbag/meetings/jan2015/presentations/SBAG12_ARM_Candidates_Chodas.pdf
[62] http://www.lpl.arizona.edu/css/
[63] http://pan-starrs.ifa.hawaii.edu/public/asteroid-threat/near-earth.html
[64] http://fallingstar.com/home.php

Asteroid miners need additional serious astronomy to do two essential things: (1) get precise orbits so that newly discovered asteroids, which are mostly small (100 m diameter or less) are not lost again, and (2) get a first order clue to the composition of the asteroids of interest. (I.e. are they stony, carbonaceous or metallic?). Among optically discovered asteroids 85% are found to be stony and so not too interesting for asteroid miners as they are low in water or precious metal content [14]. Hence just being able to rule out the stony asteroids will increase the pay-off from proximity prospecting probes seven-fold. Professional astronomers using a 4-meter class telescope with optimized instrumentation could do both jobs. The venture could be undertaken as a government, philanthropic or commercial project [22].

(4) The challenges of working with a large "uncooperative" mass in zero-g are not trivial, and matter greatly to would-be asteroid miners. These same challenges will be addressed by NASA's Asteroid Redirect Mission (ARM)[66], which is planned to return a multi-ton boulder from an asteroid to a lunar orbit by the mid-2020s. Once returned, commercial companies may be invited to test mining techniques on a small scale on this boulder. So by around 2030 there should be established beneficiation (ore-concentration) techniques that have been tested in space.

4.3 LUNAR RESOURCES

Given that no-one has been to the Moon for over 40 years, lunar mining may seem unlikely. Here too developments are promising for the use of lunar resources.

Multiple spacecraft have mapped out the Moon's composition – its resources – over the past decade [13]. The resulting maps show a varied, highly clumped, distribution of interesting materials[67]. Some parts of the Moon are far more useful than others. The permanently dark craters at the lunar poles are foremost in this list, as they seem to have preserved volatiles, including water deposited there by asteroids and comets eons ago. As a commercial mining site the Moon has advantages over asteroids. Having some gravity can be helpful in mining and refining operations. Also the Moon is always in essentially the same location, and so is permanently accessible with short journey times of a few days. The time cost of money makes flexible access and short transit times important to a business.

---

[65] http://neocam.ipac.caltech.edu;
http://www.lpi.usra.edu/sbag/meetings/jan2016/presentations/Mainzer.pdf
[66] https://www.nasa.gov/mission_pages/asteroids/initiative/index.html
[67] Crawford also gives a convincing critique of the case for $^3$He mining on the Moon.

Shackleton Energy[68] has ambitious plans to mine water commercially from the South Pole permanently dark craters. There are other permanently dark craters, but Shackleton crater at the South Pole of the Moon is special in that there are a series of thin ridges and crater rims just a few kilometers away that are in near permanent sunlight. These poetically named "Peaks of Eternal Light" [23] could provide a cheap continuous power source for mining the water. The solar panels would have to be mounted several meters above ground for truly uninterrupted power. The Peaks of Eternal Light cover less than 1 sq. km, and so are a scarce resource that will be a source of potential conflict if lunar mining becomes a real industry [24].

So the first commercial extraction of lunar resources is looking plausible by 2035.

4.4 LUNAR TELESCOPES

Traditionally the main band in which a lunar observatory has an advantage is low frequency radio telescope on the far-side, where it would be shielded from terrestrial interference. A large version of this telescope with long baselines to obtain good angular resolution would be straightforward to construct if we have the capability to mine the lunar surface. We would need to take care than any GPS system or comsats in lunar orbit do not produce interference for this telescope.

Lunar near-side short wavelength (sub-mm) telescopes working in concert with the very long baseline interferometer (VLBI) Event Horizon Telescope (EHT)[69] would allow the EHT to image the event horizons of ~30 supermassive black holes, rather than the 2 that are accessible using Earth diameter baselines. Implementing VLBI requires knowing the separations of the telescopes to a fraction of a wavelength. The Moon provides a far more stable baseline than any spacecraft can currently offer.

A newly emerged possibility is that of large lunar mid- to far-infrared telescopes. The JWST primary mirror is cooled to ~50 K which allows it to be sensitive out to 24 micron wavelengths[70]. Designing a system to keep this 6.5 m diameter mirror at this low temperature is a challenge. In part because of this problem the HDST concept eschews wavelengths longer than about 3 microns. There are large areas inside the permanently dark craters near the lunar South Pole with stable temperatures of 25 K (about -250 C) [25]. These sites could be valuable for much larger mid- to far-infrared telescopes, including interferometers, that could work out to at least 50 micron wavelengths. On the 2035 timescale these possibilities may well become feasible. They would need to keep some distance from mining operations to avoid dust settling on their mirrors. Perhaps some permanently dark craters should be set aside for astronomy.

4.5 ON-ORBIT ASSEMBLY

Valid concerns over reliability have led the space agencies to be extremely wary of moving parts in their spacecraft. As telescopes sizes come to exceed the limits of launcher fairings however, unfolding spacecraft designs are being used. From the relatively simple extending optical benches employed in Japanese[42] and US X-ray missions[71] (for which long focal lengths are inescapable), to the complex unfolding of the JWST mirror assembly and thermal shields[72], on-orbit deployment is becoming necessary for astrophysics missions. Launcher fairings are presently about 5 meters in diameter. The limit of mirror diameter achievable by unfolding from a 5 meter fairing[4] appears to be ~12m. The later (Block 2B, late 2020s) SLS fairing is intended to reach 10 meters[73].

---

[68] http://www.shackletonenergy.com
[69] http://www.eventhorizontelescope.org
[70] The smaller (0.85m) infrared Spitzer Space Telescope[70] was initially cooled to about 5 K and was sensitive out to 160 microns.

In order to build the larger telescopes that astronomers need (Section 2) we will at some point have to go to the next step: on-orbit assembly. For example, any version of the Terrestrial Planet Imager - Interferometer involves linking several smaller telescopes with tethers or trusses (Figure 4, Section 2.3). Telescopes like these will need on-orbit assembly, using pre-built components being delivered on separate (lower cost) launchers.

Missions to the Outer Planets would benefit in either transit time or delivered mass from on-orbit mating to a fully-fueled upper stage, or re-fueling of its own upper stage (Section 4.2). Longer magnetometer booms are desirable. Larger communications antennae and solar panel arrays could also use on-orbit assembly and may be mission enabling.

Similar assembly capabilities will need to be developed for lunar or asteroid mining operations. Extracting 1000 tons of water from a 10,000 ton, or larger, asteroid is challenging. A sufficiently capable mining spacecraft, that can deal with these Olympic swimming pool volumes and ISS masses, will not be small and will almost certainly involve on-orbit assembly. The late 2020s are a plausible time horizon for testing smaller scale asteroid mining spacecraft.

4.6 ON-ORBIT CONSTRUCTION

Beyond the scale on which components can be delivered from the ground and then assembled, construction on-orbit is required. That is, manufacturing components in space from raw materials. Commercial pressure for large structures to handle mining operations at asteroids or beneficiation operations in cis-lunar space (i.e. within the Earth-Moon system) will tend to lead to the development these capabilities.

Astronomy has plenty of potential for making use of large structures in space. Some examples are occulting starshades, radio dishes, support structures for shorter wavelength telescopes, interferometer trusses (up to some scale at which free-flying station-keeping satellites are more appropriate) and, ideally, large short-wavelength optics.

Plans for in-space construction are very limited at the present. An example of what might be possible on a usefully short timescale is the "SpiderFab" concept from the company Tethers Unlimited[74]. SpiderFab would build low mass carbon-fiber trusses in space[75]. At the planned rate of 3 meters/hour several SpiderFabs working in parallel may be needed to build large (~100 m class) structures on a reasonable timescale. This effort was funded in 2015 by NASA's Small Business Innovation Research (SBIR) program[76]. The company is targeting the "early 2020s" for in-space tests, if all goes well.

---

[71] *NuSTAR*: http://www.nustar.caltech.edu/page/about
[72] https://www.youtube.com/watch?v=vpVz3UrSsE4
[73] SLS Program Mission Planners Guide Executive Overview: http://www.aiaa.org/uploadedFiles/Events/Other/Student_Competitions/SLS-MNL-201%20SLS%20Program%20Mission%20Planner%27s%20Guide%20Executive%20Overview%20Version%201%20-%20DQA.pdf

The left over tailings from asteroid mining could become a major resource on the 2035 timescale. There will be strong incentives to contain the tailings from mining activities. First because clouds of fine particles from mining must not be allowed to collect around the mining spacecraft, shorting out electronics, reducing solar power yield or jamming moving parts. Second, if allowed to be free the tailings will stretch out along the asteroid orbit creating new meteor showers for Earth but also threatening satellites in Earth orbit. So for both mining and liability reasons the tailings must be contained. Tailings are at least 90% of the mass of the asteroid. They can slowly be sent back to a high Earth orbit at small incremental cost. Once they arrive they will provide construction materials in the 10,000 ton range that would cost trillions to bring from Earth. Building large equipment in space then becomes feasible.

However the timescale for these large construction activities is at least at the 2035 limit of the horizon considered here. Space agencies could hasten the arrival of this capability by funding efforts in this direction, including in-space tests at the ISS or elsewhere.

4.7 RAPID SOLAR SYSTEM TRANSIT TIMES

Transit times for reaching space resources and returning them to where they will be used have a big effect on the return on investment (ROI) for commercial operations because of the "time is money" imperative already noted in Section 2.2.2. The need to move large masses (1000s of tons for water, 10s of tons in the alternate case of precious metals not discussed here) rapidly from asteroids at AU-class distances[77] implies more powerful rocketry than we now possess.

The ARM mission is a pathfinder in this direction. ARM (see Section 4.2) scales up SEP by a factor ~5, from ~8 kW for today's GEO communication satellites (comsats) to ~40 kW. A similar scaling from ARM to ~200 kW is being considered by NASA for Mars missions[43] (Section 4.2), and does not appear to be a major stretch. Solar- and nuclear-electric (radioisotope) propulsion at 250 kW has been studied for Mars missions [26]. Electric propulsion can move ~150 tons from LEO to Mars orbit. The journey takes 3.7 years, though, and so is for cargo only. Spiraling out of LEO to Earth-escape and down to Mars orbit takes most of that time, with the transit to Mars taking only 1.7 years. Systems assembled in high orbit, e.g. at Earth-Moon L1 and fueled from lunar or asteroid resources can avoid the spiral-out phase; rendezvous with an asteroid using only SEP or NEP will be lengthy, so hybrid systems using chemical propulsion for rendezvous may be called for.

---

[74] http://www.tethers.com/AboutTUI.html#
[75] http://www.space.com/28846-spiderfab-space-structures-incredible-technology.html
[76] http://sbir.nasa.gov
[77] 1 AU = mean Sun-Earth distance ~ 150 million km.

Based on ARM, NASA has based the "split mission" concept for journeys to Mars on a scaled up, ~200 kW, version of the ARM SEP[78]. Supplies would be sent to Mars orbit, or to Phobos or Deimos, using SEP in advance of the human crew. At the moment these plans use the Space Launch System (SLS) to launch the cargo mission, but a shift to on-orbit assembly and fueling is quite possible if those capabilities progress rapidly. Steps toward developing the advanced SEP for such missions will likely have begun by the mid-2020s and will facilitate both asteroid mining and planetary science.

For Solar System science much smaller payloads of only a few tons need to be delivered. But high power propulsion is essential. With only 15 kW of solar or radioisotope electric propulsion a Uranus orbiter mission still has a daunting 13-year journey out [27]. Clearly a 200 kW-class electric propulsion mission would arrive sooner, but studies are needed to know how just much.

A more distant prospect is to use nuclear fission-powered rockets [28]. In a way space is the ideal place to use nuclear reactors; interplanetary space is already heavily irradiated by the solar wind and by Galactic cosmic rays, so our activities will not have any significant effect on that environment. Until asteroid mining becomes a major industry with political clout the substantial regulatory hurdles against using nuclear reactors in space are not likely to be overcome.

NASA is out in front on faster transit with its ARM and Mars plans. The need to limit Mars-bound astronauts' exposure to radiation, primarily against Galactic cosmic rays[79], will keep the issue high on NASA's list of priorities. If encouraging the harnessing of space resources were given to NASA as a priority then they could do even more.

It seems safe to say that this class of Solar System rapid transit will arrive no earlier than the end of the time horizon being considered in this paper.

4.8. THE POTENTIAL STATE OF PLAY IN 2030

An overview of the possible developments in commercial space in the 2020-2030 decade that have been discussed in this section is given in Table 3.

The effect that these developments will have on our in-space capabilities and so on astronomy and planetary science are also listed there. On this timescale there is of course great uncertainty about how far these gains will have progressed. The consequences for science will most likely be just beginning rather than fully realized.

---

[78] Gates, M., 2015, http://www.lpi.usra.edu/sbag/meetings/jul2014/presentations/0900_Wed_Gates_ARM_activities.pdf
[79] Wilson, J.W. et al. 1998: http://www.cs.odu.edu/~mln/ltrs-pdfs/NASA-98-mrsfm-jww.pdf

Table 3: Potential Commercial Space Activities in 2030 and their implications for astronomy and planetary science.

| COMMERCIAL SPACE ACTIVITY | EFFECT | VALUE TO ASTRONOMY, PLANETARY SCIENCE |
|---|---|---|
| Private Space Station in HEO | LEO-HEO transport | Servicing in ES-L2<br>Fueling capability<br>On-orbit assembly |
| Lunar Resources | Large masses of water in HEO<br>Major operational capability on lunar surface. | Fueling interplanetary/Mars missions in HEO<br>Lunar far-side radio telescopes.<br>Far-infrared lunar telescopes. |
| Asteroid Resources | Assembly of large mining equipment in HEO<br>High power SEP/NEP<br>Large masses of water in HEO<br>Cheap construction material from tailings | Fuel for interplanetary/Mars mission in HEO<br>Short journey times to Outer Solar System<br>Large masses to Inner Solar System, including Mars<br>Large on-orbit construction |

5. NON-TECHNICAL ISSUES

Not every issue facing astronomers and planetary scientists is technical. A few of the non-technical issues are discussed briefly below. They are not easy to solve, but need to be stated so that we can begin to consider how to address them.

5.1 TIMELINES

All extrapolations such as the ones above are fraught with uncertainty. Especially in space ventures all dates have a tendency to "slip to the right", sometimes by 5-10 years.

Nevertheless section 3 seems to show that it would not be prudent to ignore the potential near-term effects to about 2020 that commercial space resource exploitation could have on astronomy and planetary science.

The longer term changes to about 2030 discussed in section 4 are evidently more speculative, but bear following carefully as their potential impact is greater than the near-term changes.

5.2 ADAPTATION

Once the technical solutions exist to build lower cost missions there will remain the significant issue of convincing space engineers to adapt to these approaches. Decades of experience focused on minimizing mass have led to a large knowledge base for this approach. Abandoning this expertise in building finely crafted bespoke spacecraft in favor of simpler, low complexity, batch produced spacecraft will not be done lightly. In order to transition to the new paradigm the profession will need some incentive. Materially, the incentive will the danger of going out of business if any other satellite builder adopts the approach. Intellectually and emotionally the incentive will be that, instead of repeatedly solving essentially the same problem, space engineers will at last be able to move on to new problems. A large number of these are implied in this study: LEO-HEO transport; HEO laboratories and habitats; space mining equipment, both lunar and asteroid; lunar surface to orbit transport; high power SEP/NEP rockets; fuel depots and on-orbit refueling. These are surely more interesting, and potentially more profitable, problems to solve.

Achieving low cost missions involves using a novel cost model. Adopting a new cost model is a big and risky step when a flagship mission is at stake. NASA and the other space agencies will rightly demand convincing evidence that this new approach is both truly cheaper and will deliver reliable spacecraft. The probable path is that commercial satellites will first demonstrate that the approach works, and the industry will have suitable spacecraft ready for purchase. Space agencies will likely then try out the approach on medium-class missions first, before committing to building flagship missions this way. This approach would delay the gains for astronomy and planetary science, but would make them more certain in the longer run.

5.3 TEMPTATION: AVOIDING THE 'ONE BIG MISSION'

For any scientist the temptation will be strong to use any savings on launch and spacecraft to create an enhanced science payload: "Payloads grow to fill the budget available", you might say. This is the easy approach. Discipline will be needed to restrain the scientists from spending all the gains on more payload rather than on a second flagship. This may be a discipline mandated by the space agencies, or it may be a self-imposed discipline based on a community consensus that more missions are better than one.

A large number of missions are better than one giant mission. (Each one, of course, still needs to be a big step over its predecessor.) There are several arguments for this position. First, a single large mission is a risk to the program. If it fails, there is no back-up. Second, the science of one mission cannot span the requirements of the field as a whole and the synergy we now have will thus be lost (Section 2.2). Third, but equally important, is that a diversity of missions is also a good for the intellectual vitality of the field. If all astronomers use just one telescope, this will limit the breadth of their scientific imagination. Scientists dependent on being granted time on a single telescope feel a pressure to go along with current fashions in what questions and approaches are important. Time assignment committees tend to be made up of those who were successful in earlier rounds, reinforcing this narrowing trend. These effects will tend to restrict creative new approaches. Exoplanets are an example of how the ability of small teams of astronomers to take a different path led to an explosion of new science utterly unanticipated by the large majority of astronomers [29].

Sociologically it is harder to push for multiple diverse missions than for one giant mission. A giant mission proposal tends to accrete a large number of scientists and a single large community speaks louder than several smaller ones.

Politically it is also hard to "sell" a list of excellent missions both to the space agencies and to the governments that will pay for them. Saying these are the "two best" missions has intrinsically less force than saying this is "the best" mission.

The astronomy and planetary science communities should actively consider and debate these issues.

6. CONCLUSIONS

The timeline for the next large space observatories reaches out to 2030 – 2035. For the past 3 decades astronomers and planetary scientists have not had to think much about how space technology would change within their planning horizon. However, this time around enormous improvements in space infrastructure capabilities and, especially, costs are likely and can be anticipated in reasonable detail. Thanks to a new commercial focus for space ventures, substantially cheaper launchers will be the norm by 2020. Lower cost launch will lead to more massive and much cheaper, "off-the-rack" spacecraft. Also by 2020 human activity in low Earth orbit is likely to grow with privately run passenger vehicles both for tourism and to service commercially-run milli-gravity research laboratories. Biotech companies are likely to be major customers of these laboratories. On-orbit servicing of observatories can then return at a reasonable cost. Lunar landers and "interplanetary cubesats" can allow a new approach to planetary science.

The result of these changes will be that flagship missions could be proposed to the US 2020/2021 Astronomy and Planetary Science decadal reviews that cost a fraction of their present multi-billion dollar price tags. We can then have a full complement of well-matched "Greater Observatories", spanning the whole of the electromagnetic, and now gravitational wave, spectrum. To achieve this result we must consider how to resist letting our aspirations expand to telescopes that cost as much as at present. This is, admittedly, a continual struggle.

Of course, the timeline I have presented may be too optimistic. Conversely, profit is a powerful motivator. Events may move swiftly if large profits are anticipated or realized. Scientists should be prepared.

Astronomers and planetary scientists would be well advised to include these developments in our planning, at least as contingencies. We should also urge our funding agencies to actively engage with the consequences of these developments and to help make them happen. "*Only begin!*", as Goethe did not really say[80].

---

[80] The English lines attributed to Goethe in the Prelude in the Theater of Faust are: "Only begin! What you can do, or dream you can, begin it; Boldness has genius, power and magic in it; Only engage and then the mind grows heated; Begin, and then the work will be completed". (e.g. https://books.google.com/books?id=gwoWAAAAYAAJ&pg=PA108&lpg=PA108&dq=%22only+begin%22+goethe&source=bl&ots=K16fcQb5z1&sig=M8iOxFcSYqD1dg5_Jjs2j4PoFTI&hl=en&sa=X&ved=0ahUKEwjuyaGTo9LLAhXHHD4KHbYuDIQQ6AEIJzAC#v=onepage&q=%22only%20begin%22%20goethe&f=false). This is wonderful, but is at best a very free translation (http://german.about.com/library/blgermyth12.htm).

## Acknowledgements

I thank Jonathan McDowell, who is an inexhaustible resource for detailed, reliable space information and wisdom, and Jeff Hoffman for providing a quote about servicing Hubble. I thank the Aspen Center for Physics funded by NSF grant # 1066293 for their hospitality while this paper was initiated.